\documentclass[twocolumn,secnumarabic,amssymb,amsmath, nobibnotes, aps, prd, letterpaper]{revtex4}
\usepackage{graphicx}
\usepackage{bm}

\begin{document}

\title{Soliton absorption spectroscopy}
\author{V. L. Kalashnikov}
\email{kalashnikov@tuwien.ac.at}
\homepage{http://info.tuwien.ac.at/kalashnikov}
\affiliation{Institut f\"{u}r Photonik, TU Wien, Gusshausstr. 27/387,
1040 Vienna, Austria}
\author{E. Sorokin}
\email{sorokin@tuwien.ac.at}
\affiliation{Institut f\"{u}r Photonik, TU Wien, Gusshausstr. 27/387,
1040 Vienna, Austria}

\begin{abstract}
We analyze optical soliton propagation in the presence of weak
absorption lines with much narrower linewidths as compared to the
soliton spectrum width using the novel perturbation analysis
technique based on an integral representation in the spectral
domain. The stable soliton acquires spectral modulation that follows
the associated index of refraction of the absorber. The model can be
applied to ordinary soliton propagation and to an
absorber inside a passively modelocked laser. In the latter case, a
comparison with water vapor absorption in a femtosecond Cr:ZnSe
laser yields a very good agreement with experiment. Compared to the
conventional absorption measurement in a cell of the same length,
the signal is increased by an order of magnitude. The obtained
analytical expressions allow further improving of the sensitivity
and spectroscopic accuracy making the soliton absorption
spectroscopy a promising novel measurement technique.
\end{abstract}
\maketitle

\section{introduction}

Light sources based on femtosecond pulse oscillators have now
become widely used tools for ultrashort studies, optical metrology,
and spectroscopy. Such sources combine broad smooth spectra with
diffraction-limited brightness, which is especially important for
high-sensitivity spectroscopic applications. Advances in near- and
mid-infrared femtosecond oscillators made possible operation in the
wavelength ranges of strong molecular absorption, allowing direct
measurement of important molecular gases with high resolution and
good signal-to-noise ratio \cite{extra}. At the same time, it was
observed that such oscillators behave quite differently, when the
absorbing gas fills the laser cavity or introduced after the output
mirror \cite{EP-exp,EP-theor}. The issue has become especially important
with introduction of the mid-IR femtosecond oscillators such as
Cr:ZnSe \cite{CZS-1}, which operate in the 2--3 $\mu$m wavelength
region with strong atmospheric absorption.

\begin{figure}
\includegraphics[width=\columnwidth]{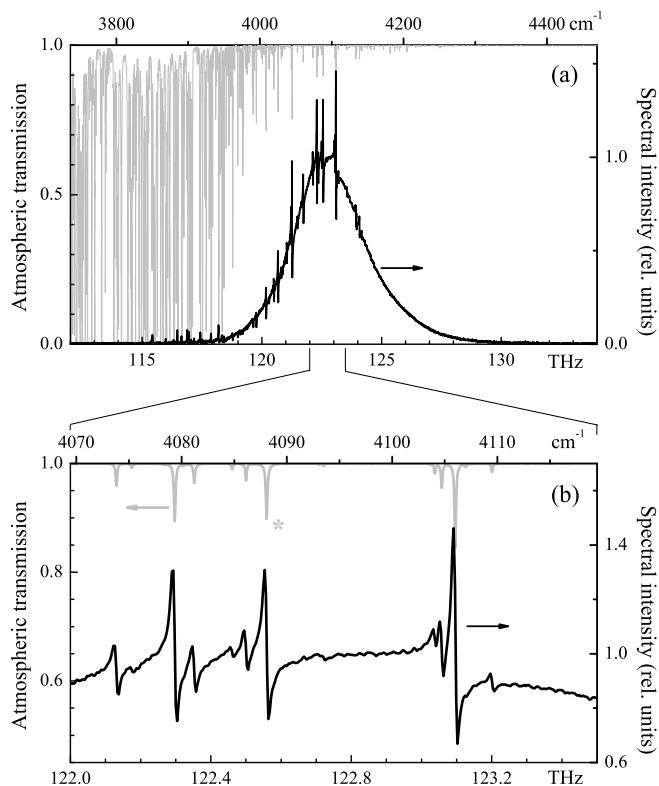}
\caption{\label{fig:CrZnSe} Output spectrum of a
100-fs Cr:ZnSe oscillator (black solid line) when operated at open air. The
atmospheric transmission (gray) is calculated from HITRAN  database
\cite{HITRAN} and corresponds to a full round-trip. The lower graph
\emph{(b)} shows the expanded central part of the spectrum. Asterisk denotes the absorption line,
which is used for quantitative evaluation in the last section.}
\end{figure}

As an example, Figure \ref{fig:CrZnSe} presents a typical spectrum
of a Cr:ZnSe femtosecond oscillator, operating at normal atmospheric
conditions. It is clearly seen, that the
pulse spectrum acquires strong modulation features which resemble
the dispersion signatures of the atmospheric lines. Being undesirable
for some applications, such spectral modulation might at the same
time open up interesting opportunity of intracavity absorption spectroscopy.
Compared with the traditional intracavity laser absorption spectroscopy
\cite{ICLAS-gen,ICLAS-CZS} based on transient processes, this approach
would have an advantage of being a well-quantified steady-state technique,
that can be immediately coupled to frequency combs and optical frequency
standards for extreme accuracy and resolution.

In this paper, we present a numerical and analytical treatment of
the effect of a narrowband absorption on a femtosecond pulse,
considered as a dissipative soliton. Such a treatment covers both,
passively modelocked ultrashort pulse oscillators with intracavity
absorbers, and soliton propagation in fibers with impurities. The
theoretical results are compared with the experiment for a
femtosecond Cr:ZnSe oscillator operating at normal atmospheric
conditions. We prove that the spectral modulation imposed by a
narrowband absorption indeed accurately follows the associated index
of refraction when the absorber linewidth is sufficiently narrow.

\section{The model}

Our approach is based on the treatment of an ultrashort pulse
as one-dimensional dissipative soliton of the nonlinear complex
Ginzburg-Landau equation (CGLE) \cite{Haus-apm,nail}. This equation
has such a wide horizon of application that the concept of {}``the
world of the Ginzburg-Landau equation'' has become broadly
established \cite{kramer}. In particular, such a model describes a
pulse with the duration $T_{0}$ inside an oscillator or propagating
along a nonlinear fiber.

To obey the CGLE, the electromagnetic field with the amplitude $A(z,t)$
should satisfy the slowly-varying amplitude approximation, provided
by the relation $\omega_{0}\gg1/T_{0}$, where $\omega_{0}$ is the
field carrier frequency, $t$ is the local time, and $z$ is the propagation
coordinate. This approximation is well satisfied even for pulses of
nearly single optical cycle duration \cite{TB-FK}. When additionally we
can neglect the field variation along the cavity round-trip or the
variation of material parameters along a fiber, as well as the contribution
of higher-order dispersions, the amplitude dynamics can be described
on the basis of the generalized CGLE \cite{nail,kaertner,nail-2,akhmanov}
\begin{widetext}
\begin{equation}
\frac{\partial A(z,t)}{\partial z}=\left\{ -\sigma+\hat{\Sigma}[P(z,t)]-
i\gamma P(z,t)\right\} A(z,t)+\left(\alpha+i\frac{\beta_{2}}{2}\right)
\frac{\partial^{2}}{\partial t^{2}}A(z,t)+\hat{\Gamma}\left[A(z,t)\right],\label{eq:MainEQ}
\end{equation}
\end{widetext}
\noindent where $P\equiv\left|A\right|^{2}$ is the instant field
power and $\alpha$ is the inverse gain bandwidth squared. The nonlinear
terms in Eq. (\ref{eq:MainEQ}) describe i) saturable self-amplitude
modulation (SAM) with nonlinear gain defined by the nonlinear operator
$\hat{\Sigma}$, and ii) self-phase modulation (SPM), defined by the
parameter $\gamma$. For a laser oscillator, $\gamma=4\pi nn_{2}l_{cryst}/\lambda_{0}A_{eff}$.
Here $\lambda_{0}$ is the wavelength; $n$ and $n_{2}$ are the linear
and nonlinear refractive indexes of an active medium, respectively;
$l_{cryst}$ is the length of the active medium; $A_{eff}=\pi w^{2}$
is the effective area of a Gaussian mode with the radius $w$ inside
the active medium. The propagation coordinate $z$ is naturally normalized
to the cavity length, i.e. $z$ becomes the cavity round-trip number.
For a fiber propagation, $\gamma=2\pi nn_{2}/\lambda_{0}A_{eff}$,
where $n$ and $n_{2}$ are the linear and nonlinear refractive indexes
of a fiber, respectively, and $A_{eff}$ is the effective mode area
of the fiber \cite{agrawal}. Finally, $\beta_{2}$ is the round-trip
net group delay dispersion (GDD) for an oscillator or the group velocity
dispersion parameter for a fiber with $\beta_{2}<0$ corresponding
to anomalous dispersion.

The typical explicit expressions for $\hat{\Sigma}[P]$ in the case,
when the SAM response is instantaneous, are i) $\hat{\Sigma}[P]=\kappa P$
(cubic nonlinear gain), ii) $\hat{\Sigma}[P]=\kappa\left(P-\zeta P^{2}\right)$
(cubic-quintic nonlinear gain), and iii) $\hat{\Sigma}[P]=\kappa P/\left(1+\zeta P\right)$
(perfectly saturable nonlinear gain) \cite{nail,nonkerr}. The second
case corresponds to an oscillator mode-locked by the Kerr-lensing
\cite{Haus-apm}. The third case represents, for instance,
a response of a semiconductor saturable absorber, when $T_{0}$ exceeds its excitation
relaxation time \cite{sesam}. However, if the latter condition is
not satisfied, one has to add an ordinary differential equation for
the SAM and Eq. (\ref{eq:MainEQ}) becomes an integro-differential
equation (see below).

The $\sigma$-term is the saturated net-loss at the carrier frequency
$\omega_{0}$, which is the reference frequency in the model. This
term is energy-dependent: the pulse energy
$E(z)\equiv\intop_{-\infty}^{\infty}P(z,t)dt$ can be expanded in the
vicinity of threshold value $\sigma=0$ as
$\sigma\approx\delta\left(E/E^{*}-1\right)$ \cite{apb}, where
$\delta=\ell^{2}/g_{0}$ ($\ell$ is the frequency-independent loss
and $g_{0}$ is the small-signal gain, both for the round-trip) and
$E^{*}$ is the round-trip continuous-wave energy equal to the
average power multiplied by the cavity period.

The operator $\hat{\Gamma}$ describes an effect of the frequency-dependent
losses, which can be attributed to an absorption within the dissipative
soliton spectrum. That can be caused, for instance, by the gases filling
an oscillator cavity or the fiber impurities for a fiber oscillator.
Within the framework of this study, we neglect the effects of loss
saturation and let $\hat{\Gamma}$ be linear with respect
to $A\left(z,t\right)$. The expression for $\hat{\Gamma}\left[A(z,t)\right]$
is more convenient to describe in the Fourier domain, $\tilde{A}(z,\omega)$
being the Fourier image of $A(z,t)$. If the losses result from the
$l$ independent homogeneously broadened lines centered at $\omega_{l}$
(relative to $\omega_{0}$) with linewidths $\Omega_{l}$ and absorption
coefficients $\epsilon_{l}<0$, then the action of
operator $\hat{\Gamma}$ can be written down in the form of a superposition
of causal Lorentz profiles \cite{butilkin,lorentz}

\begin{equation}
\hat{\Gamma}\left[\tilde{A}\right]=\left(\sum_{l}\frac{\epsilon_{l}}
{1+i(\omega-\omega_{l})/\Omega_{l}}\right)\tilde{A}(z,\omega).\label{eq:lorentz}
\end{equation}

\noindent In the more general case the causal Voigt profile has to
be used for $\hat{\Gamma}\left[\tilde{A}\right]$ \cite{cVoigt}.
Causality of the complex profile of Eq. (\ref{eq:lorentz}) demonstrates
itself in the time domain, where one has

\begin{equation}
\hat{\Gamma}\left[A\left(z,t\right)\right]\propto\sum_{l}\epsilon_{l}
\Omega_{l}\intop_{-\infty}^{t}e^{-\left(\Omega_{l}-i\omega_{l}\right)
\left(t-t'\right)}A\left(z,t'\right)dt'.\label{eq:time}
\end{equation}

The conventional analysis of perturbed soliton propagation includes
approximation of the effective group delay dispersion of the
perturbation as Taylor series
$\beta^{\prime}(\omega)=\beta_{2}^{\prime}(\omega-\omega_{0})^{2}/2+\beta_{3}^{\prime}(\omega-\omega_{0})^{3}/6,\dots$
assuming that the additional terms
$\beta_{2}^{\prime},\beta_{3}^{\prime},\dots$ are sufficiently
small. This approach is absolutely not applicable in our case,
because the dispersion, associated with a narrow linewidth absorber
can be extremely large. For example, an atmospheric line with a
typical width $\Omega=3$GHz and peak absorption of only $10^{-3}$
produces the group delay dispersion modulation of
$\beta_{2}^{\prime}=\pm0.9$ ns$^{2}$, far exceeding the typical
intracavity values of $\beta_{2}\sim10^{2}..10^{4}$ fs$^{2}$.
Moreover, decreasing the linewidth $\Omega$ (and thus reducing the
overall absorption of the line) causes the group delay dispersion
term to diverge as $\beta_{2}^{\prime}\propto\Omega^{-2}$.

In the following we shall therefore start with a numerical analysis
to establish the applicability and stability of the model, and then
present a novel analysis technique based on integral representation
in the spectral domain.

\section{Numerical analysis}

Without introducing any additional assumptions, we have solved the
Eqs. (\ref{eq:MainEQ},\ref{eq:lorentz}) numerically by the
symmetrized split-step Fourier method. To provide high spectral
resolution, the simulation local time window contains $2^{22}$
points with the mesh interval 2.5 fs. The simulation parameters for
the cubic-quintic version of Eq. (\ref{eq:MainEQ}) are presented in
Table \ref{tab:SimulParams}. The GDD parameter $\beta_{2}=-$1600
fs$^{2}$ provides a stable single pulse with FWHM $\approx$100 fs.
The single low-power seeding pulse converges to a steady-state
solution during $z\approx$5000.

\begin{table}
\begin{centering}
\begin{tabular}{|c|c|c|c|c|c|c|}
\hline $E^{*}$ & $\gamma$ & $\alpha$ & $\kappa$ & $\zeta$ & $\delta$
& $\beta_{2}$\tabularnewline \hline \hline 20 nJ & 2.5 MW$^{-1}$ &
16 fs$^{2}$ & 0.02$\gamma$ & 0.2$\gamma$ & 0.03 & -1600
fs$^{2}$\tabularnewline \hline
\end{tabular}
\par\end{centering}

\caption{\label{tab:SimulParams}Laser simulation parameters. The
numbers correspond to a Cr:ZnSe femtosecond oscillator of Fig.
\ref{fig:CrZnSe} with $l_{cryst}=$0.4 cm, $w=$80 $\mu$m,
$\lambda_{0}=$2.5 $\mu$m, $n=2.44$, $n_{2}=10^{-14}$ cm$^{2}/$W,
$\ell=$0.075, $g_{0}=$2.5$\ell$.}

\end{table}

\begin{figure}
\includegraphics[clip,width=\columnwidth]{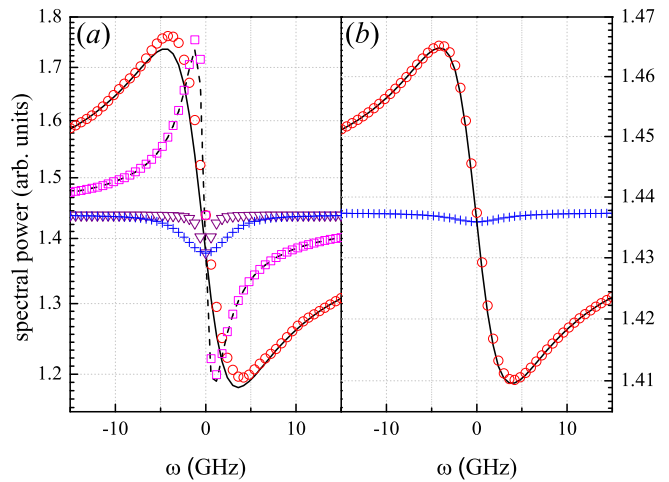}
\caption{\label{fig:Single}(Color online) Central parts of the
dissipative soliton spectra in an oscillator with the single
absorption line centered at $\omega=$0. (\emph{a})
$\epsilon=-0.05$; $\Omega=4$ GHz (solid curve, open circles
and crosses) and 1 GHz (dashed curve, open squares and triangles).
(\emph{b}) $\epsilon=-0.005$; $\Omega=4$ GHz. The solid and
dashed curves correspond to the contribution of $\hat{\Gamma}$
described by Eq. (\ref{eq:lorentz}). Open circles and squares
describe the separate contribution of
$\Im\left(\hat{\Gamma}\right)$. Crosses and triangles describe
the separate contribution of $\Re\left(\hat{\Gamma}\right)$.}
\end{figure}

The pulse propagation within a linear medium (e.g. an absorbing gas
outside an oscillator, a passive fiber containing some impurities,
or microstructured fiber filled with a gas) is described by Eqs. (\ref{eq:MainEQ},\ref{eq:lorentz})
with zero $\alpha$, $\gamma$, $\hat{\Sigma}$, and the initial $A\left(0,t\right)$
corresponding to output oscillator pulse. The obvious effect of the
absorption lines on a pulse spectrum are the dips at $\omega_{l}$
(Fig. \ref{fig:Single}a), that simply follow the Beer's law. This
regime allows using the ultrashort pulse for conventional absorption
spectroscopy \cite{extra}. The nonzero real part of an absorber permittivity
(i.e. $\Im\left(\hat{\Gamma}\right)\neq$0) does significantly change
the pulse in time domain \cite{JapsLin}, but does not alter the spectrum.
The pulse spectrum reveals only imaginary part of an absorber permittivity
(i.e. $\Re\left(\hat{\Gamma}\right)\neq$0).

\begin{figure}
\includegraphics[clip,width=\columnwidth]{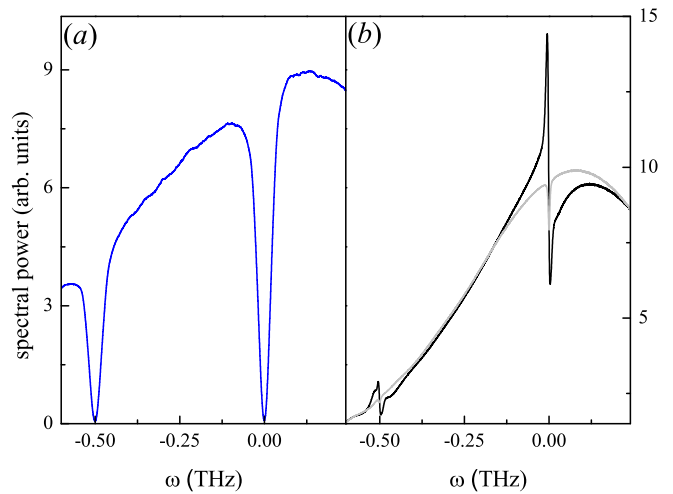}
\caption{\label{fig:Soliton}(Color online) Part of the pulse
spectrum after: (\emph{a}) linear propagation for 25 dispersion
lengths inside a fiber with two absorption lines and (\emph{b})
perturbed soliton propagation for 100 dispersion lengths. Gray curve
in (\emph{b}) corresponds to the contribution of only
$\Re(\hat{\Gamma})$ in Eq. (\ref{eq:lorentz}), black curve
corresponds to the contribution of the complex profile
(\ref{eq:lorentz}). $\epsilon_{1}=\epsilon_{2}=-0.1$,
$\Omega_{1}=\Omega_{2}=40$ GHz. The initial pulse profile is
$A\left(t\right)=A_{0}\mathrm{sech}\left(t/T_{0}\right)$, where
$T_{0}=$57 fs. }
\end{figure}

Introducing the nonzero SPM coefficient
$\gamma=|\beta_{2}|/A_{0}^{2}T_{0}^{2}$ with zero $\alpha$ and
$\hat{\Sigma}$ transforms Eq. (\ref{eq:MainEQ}) to a perturbed
nonlinear Schr\"{o}dinger equation and results in the true perturbed
soliton propagation. In this case, as shown in Fig.
\ref{fig:Soliton}b, the situation becomes dramatically different.
Besides the dips in the spectrum shown by the gray curve
corresponding to a contribution of $\Re(\hat{\Gamma})$ only, there
is a pronounced contribution from the phase change induced by the
dispersion of absorption lines (solid curve in Fig.
\ref{fig:Soliton}b corresponds to the complex profile of
$\hat{\Gamma}$ in Eq. (\ref{eq:lorentz})). As a result, the spectral
profile has the sharp bends with the maximum on the low-frequency
side and the minimum on the high-frequency side of the corresponding
absorption line. At the same time, the dips in the spectrum due to
absorption are strongly suppressed. In addition to spectral
features, the soliton decays, acquires a slight shift towards the
higher frequencies, and its spectrum gets narrower due to the energy
loss.

The soliton spectrum reveals in this case the real part of the absorber
permittivity. However, the continuous change of the soliton shape
due to energy decay renders the problem as a non-steady state case.
The situation becomes different in a laser oscillator, where
pumping provides a constant energy flow to compensate the absorption
loss.

Let us consider the steady-state intra-cavity narrowband absorption
inside a passively modelocked femtosecond oscillator, where the
pulse is controlled by the SPM and the SAM, which is described by
the cubic-quintic $\hat{\Sigma}$ in Eq. (\ref{eq:MainEQ}) modeling
the Kerr-lens mode-locking mechanism \cite{kaertner}. Such an
oscillator can operate both, in the negative dispersion regime
\cite{kaertner} with an chirped-free soliton-like pulse, and in the
positive dispersion regime \cite{apb}, where the propagating pulse
acquires strong positive chirp. In this study, we consider only the
negative dispersion regime, the positive dispersion regime will be a
subject of following studies.

\begin{figure}
\includegraphics[clip,width=\columnwidth]{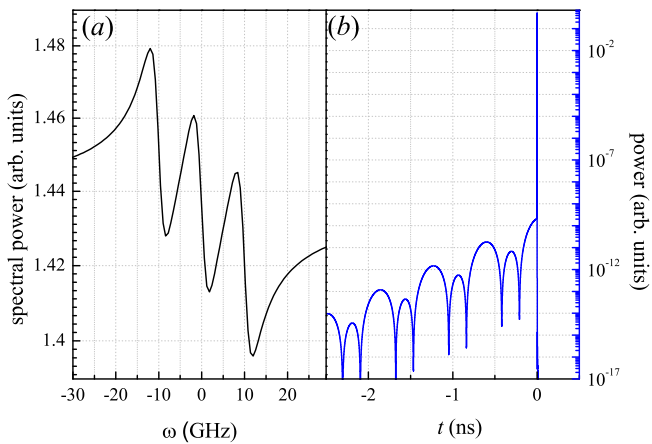}
\caption{\label{fig:Osci}(Color online) Dissipative soliton in an
oscillator: (\emph{a}) central part of the spectrum and (\emph{b})
power $P(t)$. An oscillator is filled with an absorber described by
Eq. (\ref{eq:lorentz}) with the triplet of lines:
$\epsilon_{1}=\epsilon_{2}=\epsilon_{3}=-0.005$,
$\Omega_{1}=\Omega_{2}=\Omega_{3}=2$ GHz, and $\omega_{1}=-10$,
$\omega_{2}=0$, $\omega_{3}=10$ GHz. }
\end{figure}

The results of the simulation are shown in Figs.
\ref{fig:Osci} and \ref{fig:Single}, and they demonstrate the same dispersion-like modulation of the pulse spectrum.
 Fig. \ref{fig:Osci} demonstrates action by three narrow ($\Omega=$2
GHz) absorption lines centered at -10, 0 and 10 GHz in the neighborhood of $\omega=0$.
One can see (Fig. \ref{fig:Osci}a), that the absorption lines
 do not cause spectral dips at
$\omega_{l}$, but produce sharp bends, very much like the case of
the true perturbed Schr\"{o}dinger soliton considered before. One
can also clearly see the collective redistribution of spectral power
from higher- to lower-frequencies, which enhances local spectral
asymmetry (Fig. \ref{fig:Osci}a). Such an asymmetry suggests that
the dominating contribution to a soliton perturbation results from
the real part of an absorber permittivity, which, in particular,
causes the time-asymmetry of perturbation in the time domain. This
asymmetry is seen in time domain as a ns-long modulated exponential
precursor in Fig. \ref{fig:Osci}b.

The simulated effect of a single narrow absorption line centered at
$\omega=0$ is shown in Fig. \ref{fig:Single} for different values of
peak absorption $\epsilon$ and width $\Omega$. In Fig.
\ref{fig:Single}a, $\epsilon=-0.05$ and $\Omega=$ 4 GHz (solid
curve, open circles and crosses) and 1 GHz (dashed curve, open
squares and triangles). The solid and dashed curves demonstrate the
action of complex profile (\ref{eq:lorentz}), whereas circles
(squares) and crosses (triangles) demonstrate the separate action of
$\Im(\hat{\Gamma})$ and $\Re(\hat{\Gamma})$, respectively. One can
see, that the profile of perturbed spectrum traces that formed by
only $\Im(\hat{\Gamma})$ (i.e. it traces the real part of an
absorber permittivity). One can say, that the pure phase effect
($\Im(\hat{\Gamma})$, circles and squares in Figs.
\ref{fig:Single}a,b) strongly dominates over the pure absorption
($\Re(\hat{\Gamma})$, crosses in Figs. \ref{fig:Single}a,b), like
that for the Schr\"{o}dinger soliton. Such a domination enhances
with a lowered $\epsilon$ (Fig. \ref{fig:Single}b, crosses), however
the $\left|\epsilon\right|$ growth increases the relative
contribution of $\Re(\hat{\Gamma})$ and causes the frequency
downshift of the bend (Fig. \ref{fig:Single}a). Amplitude of the
bend traces the $\epsilon$ value, while its width is defined by
$\Omega$.

The SAM considered above is modeled by the cubic-quintic nonlinear
term $\hat{\Sigma}$ in Eq.(\ref{eq:MainEQ}). Such a SAM is typically
realized by using the self-focusing inside an active medium (Kerr-lens
modelocking). For reliable self-starting operation of mode-locked
oscillator it is often desirable to use a suitable saturable absorber
(SA), e.g. a semiconductor-based SESAM \cite{Islam,keller}. Such
an absorber can be described in the simplest case by a single-lifetime
two-level model, giving the time-dependent loss coefficient $\Lambda\left(t\right)$
as

\begin{equation}
\frac{\partial\Lambda\left(t\right)}{\partial t}=\frac{\Lambda_{0}
-\Lambda\left(t\right)}{T_{s}}-\Lambda\left(t\right)\frac{P\left(t\right)}
{J_{s}A_{eff}},\label{eq:SESAM rate eq}\end{equation}

\noindent where $\Lambda_{0}$ is the loss coefficient for a small
signal, $A_{eff}$ is the effective beam area on the SA, $T_{s}$ and
$J_{s}$ are the SA relaxation time and the saturation energy
fluency, respectively. Eq. (\ref{eq:SESAM rate eq}) supplements Eq.
(\ref{eq:MainEQ}) and the SAM term $\hat{\Sigma}$ in the latter has
to be replaced by $\left(\Lambda_{0}-\Lambda\left(t\right)\right)$.
When the pulse width is longer than the SA relaxation time, one can
replace Eq. (\ref{eq:SESAM rate eq}) by its adiabatic solution so
that

\begin{equation}
\hat{\Sigma}=\dfrac{\eta_{0}\xi P\left(t\right)}{1+\xi P\left(t\right)},
\label{eq:SESAM saturation}\end{equation}

\noindent where $\xi\equiv T_{s}/J_{s}S$ is the inverse saturation
power.

We have simulated Eqs.
(\ref{eq:MainEQ},\ref{eq:lorentz},\ref{eq:SESAM rate eq}) in the
case of $J_{s}=$$50$ $\mu$ J/cm$^{2}$ and $T_{s}=$0.5 ps, that
correspond to the measurement in Fig. \ref{fig:CrZnSe}. Two cases
have been considered: weak focusing ($A_{eff}$=4000 $\mu$m$^{2}$ or
saturation energy $E_{s}$=2 nJ) and hard focusing ($A_{eff}$=1000
$\mu$m$^{2}$ or saturation energy $E_{s}$=0.5 nJ). We also considered Eq.
(\ref{eq:SESAM saturation}) for the same peak saturation level as
weakly focused SA (i.e. $\xi^{-1}$= 4 kW). In the latter case the SA effectively becomes instantaneous, and
the perturbed soliton spectrum is the same as for the Kerr-lens
modelocking, i.e. as for the cubic-quintic $\hat{\Sigma}$. When the
saturation energy is sufficiently large, there is no difference
between the models expressed by Eqs. (\ref{eq:SESAM rate eq}) and
(\ref{eq:SESAM saturation}). The effect of a narrow absorption line
is similar to that of the soliton of the cubic-quintic Eq.
(\ref{eq:MainEQ}). Decrease of the saturation energy $E_{s}$ causes
down-shift of the pulse spectrum as a whole, but the narrow bend on
the soliton spectrum reproduces the real part of the absorber
permittivity. One can thus conclude that the type of the SAM is
irrelevant for an effect of the narrowband absorption lines on a
dissipative soliton spectrum.

Another important conclusion from the numerical
simulations is the demonstrated stability of the dissipative soliton
against perturbations induced by narrowband absorption. In the
following analytical treatment we shall therefore omit the stability
analysis.

\section{Perturbative analysis of soliton spectrum}

To study the transformation of dissipative soliton spectrum under action of narrow absorption lines, we apply the perturbation method \cite{nail,maple}. Since the basic features already become apparent
for the perturbed Schr\"{o}dinger soliton and do not depend on SAM
details, we shall consider the simplest case of the cubic nonlinear
gain $\hat{\Sigma}[P]=\kappa P$. The unperturbed solitonic
chirp-free solution of such reduced equation with $\hat{\Gamma}=0$ is
$a\left(z,t\right)=A_{0}\mathrm{sech}\left(t/T_{0}\right)\exp\left[i\phi\left(t\right)+iqz\right]$
with $d\phi/dt=\varpi$. The unperturbed soliton parameters are
\cite{maple}

\begin{gather}
\varpi=0,\nonumber \\
A_{0}^{2}=\frac{2\alpha}{\kappa T_{0}^{2}},\label{eq:par1}\\
q=\frac{\beta_{2}}{2T_{0}^{2}},\nonumber \\
T_{0}=\sqrt{\frac{\alpha}{\sigma}},\nonumber
\end{gather}

\noindent where the equation parameters are confined

\begin{gather}
\beta_{2}=-2\frac{\alpha\gamma}{\kappa},\label{eq:par2}\\
\sigma>0.\nonumber \end{gather}

\noindent Hence, the soliton wavenumber is $q=-\gamma\sigma/\kappa$.

Its is reasonable to treat the soliton of the reduced Eq.
(\ref{eq:MainEQ}) as the Schr\"{o}dinger soliton with the parameters
constrained by the dissipative terms $\sigma$, $\alpha$ and $\kappa$
(see Eqs. (\ref{eq:par1},\ref{eq:par2})). This implies that the equation,
which has to be linearized with respect to a small perturbation
 co-propagating with the
soliton without beating, decay or growth (i.e. having a wavenumber real and equals to $q$ \cite{nail}), is the perturbed nonlinear Schr\"{o}dinger equation

\begin{equation}
\frac{\partial A(z,t)}{\partial z}=i\frac{\beta_{2}}{2}
\frac{\partial^{2}}{\partial t^{2}}A(z,t)-i\gamma P(z,t)A(z,t)+
\hat{\Gamma}\left[A(z,t)\right].\label{eq:schrod}
\end{equation}

\noindent Linearization of the latter with respect to a perturbation
$f\left(t\right)\exp\left(iqz\right)$ results in

\begin{gather}
iqf\left(t\right)=i\frac{\beta_{2}}{2}\frac{d^{2}f\left(t\right)}{dt^{2}}
-i\gamma\left[2\left|a\left(t\right)\right|^{2}f\left(t\right)+a\left(t\right)^{2}f^{*}
\left(t\right)\right] \nonumber \\+\hat{\Gamma}\left(a+f\right).\label{eq:perturbed}
\end{gather}

\noindent  In the spectral domain, Eq. (\ref{eq:perturbed}) becomes
\begin{widetext}
\begin{equation}
\left[k\left(\omega\right)-q\right]\tilde{f}\left(\omega\right)+\frac{1}
{\pi}\intop_{-\infty}^{\infty}d\omega^{'}U\left(\omega-\omega^{'}\right)
\tilde{f}\left(\omega^{'}\right)+\frac{1}{2\pi}\intop_{-\infty}^{\infty}
d\omega^{'}U\left(\omega-\omega^{'}\right)\tilde{f}^{*}\left(\omega^{'}\right)=
S\left(\omega\right),\label{eq:eqpqrturbation}
\end{equation}
\end{widetext}
\noindent where \cite{maple}

\begin{gather}
U\left(\omega\right)=-\pi\gamma T_{0}^{2}A_{0}^{2}\omega\mathrm{csch}\left(\pi T_{0}\omega/2\right),\nonumber \\
S\left(\omega\right)=\frac{iA_{0}\pi T_{0}}{\cosh\left(\pi T_{0}\omega/2\right)}\sum_{l}\epsilon_{l}\frac{1-i\left(\omega-\omega_{l}\right)/\Omega_{l}}{1+\left(\omega-\omega_{l}\right)^{2}/\Omega^{2}},\label{eq:source}\\
k\left(\omega\right)=-\frac{\beta_{2}}{2}\omega^{2}-\sum_{l}\epsilon_{l}\frac{\left(\omega-\omega_{l}\right)/\Omega_{l}+i}{1+\left(\omega-\omega_{l}\right)^{2}/\Omega^{2}}.\nonumber \end{gather}

\noindent Here $k\left(\omega\right)$ is the frequency-dependent
complex wave number, and $S\left(\omega\right)$ is the perturbation
source term for $\hat{\Gamma}$ corresponding to Eq. (\ref{eq:lorentz}).

Further, one may assume the phase matching between the soliton and
its perturbation. This assumption in combination with the equality
$U\left(\omega\right)=U^{*}\left(\omega\right)$, which holds for the
Schr\"{o}dinger soliton, results in
$\intop_{-\infty}^{\infty}d\omega^{'}U\left(\omega-\omega^{'}\right)
\tilde{f}^{*}\left(\omega^{'}\right)=\intop_{-\infty}^{\infty}d\omega^{'}
U\left(\omega-\omega^{'}\right)\tilde{f}\left(\omega^{'}\right)$.

The equation (\ref{eq:eqpqrturbation}) for the Fourier image of perturbation is the
Fredholm equation of second kind. Its solution can be obtained by
the Neumann series method so that the iterative solution becomes
\cite{maple}
\begin{widetext}
\begin{equation}
\tilde{f}_{n}\left(\omega\right)=\frac{S\left(\omega\right)}{k\left(\omega\right)-q}
-\frac{3}{2\pi\left[k\left(\omega\right)-q\right]}\intop_{-\infty}^{\infty}d\omega^{'}
U\left(\omega-\omega^{'}\right)\tilde{f}_{n-1}\left(\omega^{'}\right),\label{eq:neumann}
\end{equation}
\end{widetext}

\noindent where $\tilde{f}_{n}\left(\omega\right)$ is the $n$-th
iteration and $\tilde{f}_{0}\left(\omega\right)=S\left(\omega\right)/\left[k\left(\omega\right)-q\right]$.

The ``phase character'' of a soliton perturbation ($i$-multiplier
in lhs. of Eq. (\ref{eq:perturbed}) and the expression for the source
term (\ref{eq:source})) demonstrate that the real part of absorber
permittivity contributes to the real part of soliton spectral amplitude.
Simultaneously, the resonant condition $k\left(\omega\right)-q=$0,
which is responsible for a dispersive wave generation caused by, for
instance, the higher-order dispersions \cite{nail}, is not reachable
in our case. The resonance can appear in case of large $\left|\epsilon\right|$, $\kappa/\gamma$, and $\Omega_{l}T_{0}$,
but such regimes are beyond the scope of this work.

\begin{figure}
\includegraphics[clip,width=\columnwidth]{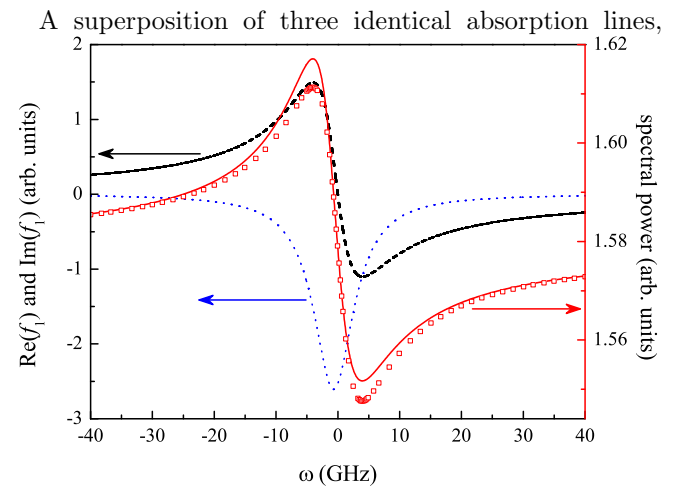}
\caption{\label{fig:f0f1}(Color online) Central part of the
dissipative soliton spectra perturbed by
$\tilde{f}_{1}\left(\omega\right)$ (solid curve) or
$\tilde{f}_{0}\left(\omega\right)$ (open squares) as well as the
profiles of $\Re\left(\tilde{f}_{1}\right)$ (dashed curve) and
$\Im\left(\tilde{f}_{1}\right)$ (dotted curve) from Eq.
(\ref{eq:neumann}). Single absorption line is centered at
$\omega=$0; $\epsilon=-0.005$; $\Omega=4$ GHz. }
\end{figure}

Eq. (\ref{eq:neumann}) can be solved numerically. Fig. \ref{fig:f0f1}
shows $\Re\left(\tilde{f}_{1}\left(\omega\right)\right)$ (dashed
curve) and $\Im\left(\tilde{f}_{1}\left(\omega\right)\right)$ (dotted
curve). One can see, that the real part of absorber permittivity defines
$\Re\left(\tilde{f}\left(\omega\right)\right)$ while the imaginary
part of absorber permittivity defines $\Im\left(\tilde{f}\left(\omega\right)\right)$.
That agrees with the simulation results and is opposite to the case
of a linear pulse propagation. One can also see a tiny frequency down-shift
$\theta$ of the of $\tilde{f}\left(\omega\right)$ minimum from $\omega_{l}$
like that in the simulations.

The pulse  spectrum  (solid curve in Fig.
\ref{fig:f0f1}), results from interference of the perturbation
with the soliton. For the chosen parameters of the absorption line
the zero-order approximation $\tilde{f}_{0}\left(\omega\right)$
(open squares) is very close to the first-order approximation (solid
curve) but is slightly down-shifted in the vicinity of the bend
maximum and minimum.

\begin{figure}[h]
\includegraphics[clip,width=\columnwidth]{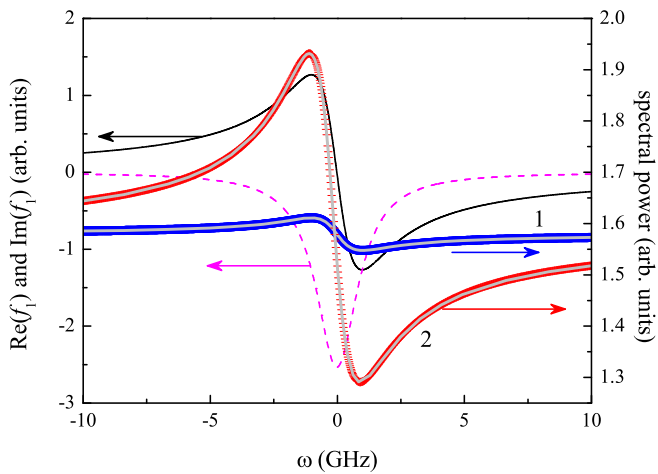}
\caption{\label{fig:f0f1-narrow}(Color online) Central part of the
dissipative soliton spectra perturbed by
$\tilde{f}_{1}\left(\omega\right)$ (open circles and crosses) or
$\tilde{f}_{0}\left(\omega\right)$ (solid gray curves) as well as
the profiles of $\Re\left(\tilde{f}_{1}\right)$ (solid black curve)
and $\Im\left(\tilde{f}_{1}\right)$ (dashed curve) from Eq.
(\ref{eq:neumann}). Single absorption line with $\Omega=1$ GHz
is centered at $\omega=$0; $\epsilon=-0.005$ (black solid
and dashed curves as well as open circles and solid gray curve 1)
and $\epsilon=-0.05$ (crosses and solid gray curve 2).  }
\end{figure}

With even narrower line width $\Omega$ of 1 GHz (Fig. \ref{fig:f0f1-narrow})
the spectral perturbation gets very close to the real part of absorber
permittivity and, simultaneously, the spectral down-shift $\theta$
(location of the $\Im\left(\tilde{f}_{1}\right)$ minimum, dashed
curve) vanishes. The $\tilde{f}_{0}\left(\omega\right)$ (gray solid
curves) now perfectly matches the $\tilde{f}_{1}\left(\omega\right)$
(open circles and crosses) within a broad range of $\epsilon$ (gray
solid curves 1 and 2 as well as circles and crosses belong to $\epsilon=$-0.005
and -0.05, respectively). The bend amplitudes are in agreement with
Eq. (\ref{eq:source}).

\begin{figure}[h]
\includegraphics[clip,width=\columnwidth]{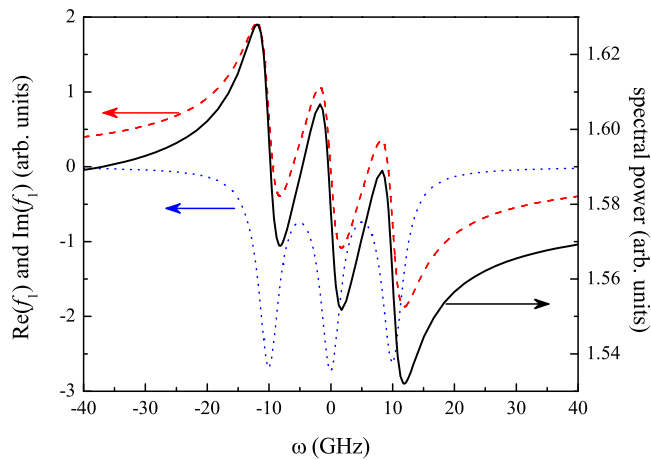}
\caption{\label{fig:3lines}(Color online) Central part of the
dissipative soliton spectra perturbed by
$\tilde{f}_{0}\left(\omega\right)$ (black solid curve) and the
profiles of $\Re\left(\tilde{f}_{0}\right)$ (dashed curve) and
$\Im\left(\tilde{f}_{0}\right)$ (dotted curve) from Eq.
(\ref{eq:neumann}). Triplet of absorption lines is centered at
$\omega_{1}=-$10, $\omega_{2}=$0, and $\omega_{3}=$10 GHz;
$\epsilon_{1}=\epsilon_{2}=\epsilon_{3}=-0.005$,
$\Omega_{1}=\Omega_{2}=\Omega_{3}=2$ GHz. }
\end{figure}

A superposition of three identical absorption lines, which corresponds
to the numerical spectra in Fig. \ref{fig:Soliton}, is shown in Fig.
\ref{fig:3lines}. One can see, that the lowest-order analytical solution
$\tilde{f}_{0}\left(\omega\right)$accurately reproduces the numerical
result. It is important, that a cumulative contribution of lines into
$k\left(\omega\right)$ does not distort a superposition contribution
of $S\left(\omega\right)$ into a soliton spectrum (see Eqs. (\ref{eq:source},\ref{eq:neumann})).
This means that the individual contribution of a single line within
a group is easily distinguishable and can be quantitatively assessed,
opening way for interesting spectroscopic applications.

As Figs. \ref{fig:f0f1} and \ref{fig:f0f1-narrow} suggest, the
zero-order approximation
$\tilde{f}_{0}\left(\omega\right)=S\left(\omega\right)/\left[k\left(\omega\right)-q\right]$
is quite accurate for a description of perturbation in the limit of
$\left|\epsilon\right|\ll1$. This allows expressing
the perturbed spectrum of an isolated line (see
(\ref{eq:source},\ref{eq:neumann})) in analytical form \cite{maple}:

\begin{widetext}
\begin{equation}
\tilde
P(\omega)\equiv\left|\tilde{a}\left(\omega\right)+\tilde{f}_{0}\left(\omega\right)\right|^2\approx
\frac{A_{0}^{2}\pi^{2}T_{0}^{2}\mathrm{sech}\left(\frac{\pi
T_{0}\omega}{2}\right)^{2}\left(\frac{\beta_{2}}{2}\omega^{2}+q\right)^{2}\left(1+\frac{\left(\omega-\omega_{l}\right)^{2}}{\Omega_{l}^{2}}\right)}
{\left[\epsilon_{l}^{2}+\frac{2\left(\omega-\omega_{l}\right)\epsilon_{l}}{\Omega_{l}}\left(\frac{\beta_{2}}{2}\omega^{2}+q\right)+\left(\frac{\beta_{2}}{2}\omega^{2}+q\right)^{2}\left(1+\frac{\left(\omega-\omega_{l}\right)^{2}}{\Omega_{l}^{2}}\right)\right]}.
\label{eq:ap1}
\end{equation}
\end{widetext}

\noindent Eq. (\ref{eq:ap1}) allows further simplification in the case of $\left|\epsilon\right|\ll1$

\begin{widetext}
\begin{equation}
\tilde P
(\omega)\approx A_{0}^{2}\pi^{2}\mathrm{sech}\left(\sqrt{\frac{\alpha}{\sigma}}\frac{\pi\omega}{2}\right)^{2}\frac{\alpha}{\sigma}\left[1+\frac{2\epsilon_{l}\kappa}{\gamma\left(\alpha\omega_{l}^{2}+\sigma\right)}\frac{\left(\omega-\omega_{l}\right)}{\left(1+\frac{\left(\omega-\omega_{l}\right)^{2}}{\Omega_{l}^{2}}\right)\Omega_{l}}\right],
\label{eq:ap2}
\end{equation}
\end{widetext}

\noindent where Eqs. (\ref{eq:par1}) and the condition $\alpha\Omega_{l}^{2}\ll1$
have been used.

Eq. (\ref{eq:ap2}) demonstrates that the spectral bend follows the real part of absorber
permittivity. Spectral downshift of bend
is the effect of $\mathcal{O}\left(\epsilon^{2}\right)$ and is not
included in (\ref{eq:ap2}). The perturbation is represented by the
term in square brackets and its relative amplitude is proportional
to $\epsilon$. Furthermore, the aspect ratio of the kink grows with
i) the increase of the relative contribution of the SAM $\kappa/\gamma$
; ii) the gain bandwidth $1/\sqrt{\alpha}$; iii) approaching of the
resonance frequency $\omega_{l}$ to the center of soliton spectrum
(but the ratio of the aspect ratio to the local soliton spectral power
increases with $\left|\omega_{l}\right|$, because the former decreases
as $\omega_{l}^{-2}$ while the latter falls faster as $\cosh\left(\pi T_{0}\omega_{l}/2\right)^{2}$);
and iv) with an approaching to the soliton stability border, which
corresponds to vanishing $\sigma$. It should be noted, that smaller
$\sigma$ entails the soliton width growth (Eq. (\ref{eq:par1})).

Since the soliton parameters are interrelated, it is instructive to
express $\sigma$ through the observable parameters such as soliton
energy $E$ or the soliton width $T_{0}$. When $\alpha\omega_{l}^{2}\ll\sigma$
(e.g. $\omega_{l}\approx$0 or/and an oscillator operates far from
the stability border $\sigma=$0), the perturbation amplitude is
inversely proportional to $\gamma\kappa E^{2}$:

\begin{equation}
\frac{2\epsilon_{l}\kappa}{\gamma\left(\alpha\omega_{l}^{2}+\sigma\right)}
\approx\frac{32\epsilon_{l}\alpha}{\gamma\kappa E^{2}}=\frac{2\epsilon_{l}\kappa T_{0}^{2}}{\gamma\alpha}
\label{eq:amp}
\end{equation}

\noindent For a fixed gain bandwidth, the amplitude scales with squared pulsewidth $T_{0}^{2}$. Ultimately,
the latter equation is equivalent to

\begin{equation}
\frac{2\epsilon_{l}\kappa}{\gamma\left(\alpha\omega_{l}^{2}+\sigma\right)}\approx-\frac{2\epsilon_{l}}{q},
\label{eq:ampshort}
\end{equation}

\noindent i.e. the relative perturbation amplitude near soliton central
frequency is the ratio of incurred loss coefficient to the soliton
wavenumber, regardless of the $z$ coordinate normalization. Therefore,
this analytical expression that has been derived for the self-consistent
oscillator, should also be valid for the case of a soliton propagation
in a long fiber when the conditions of applicability $\left|\epsilon\right| \ll 1$,
$\Omega_{l} \ll 1/T_{0}$ are met. The final form of the soliton spectrum
thus becomes

\begin{equation}
\tilde P(\omega)=\tilde
P_{0}(\omega)\left(1-\frac{2}{q}\sum_{l}\epsilon_{l}
\frac{\left(\omega-\omega_{l}\right)/\Omega_{l}}{1+\frac{\left(\omega-\omega_{l}\right)^{2}}
{\Omega_{l}^{2}}}\right), \label{eq:FinSoll}
\end{equation}

\noindent where $\tilde P_{0}(\omega)$ is the spectrum of an unperturbed soliton.

\section{Discussion}

In the analysis above we have shown that, for the case of
sufficiently sparse, narrow and weak Lorentzian absorber lines,
their spectral signatures are equivalent to the dispersion-like
modulation with a relative amplitude equal to the peak absorption
coefficient over oscillator round-trip (or nonlinear length for
passive propagation) divided by the soliton wavenumber. For
quantitative comparison with the experiment we recall equation
(\ref{eq:par1}) and express the maximum spectrum deviation of a
single line at $|\omega-\omega_{l}|=\Omega_{l}$ through observable
parameters:

\begin{equation}
\left\vert\frac{\epsilon_{l}}{q}\right\vert=\chi_{l}L\frac{T_{0}^{2}}{|\beta_{2}|}=
\chi_{l}L\frac{0.0319}{|\beta_{2}|(\Delta\nu)^{2}},
\label{eq:AmpFinal}
\end{equation}

\noindent where $\chi_{l}=2\epsilon_{l}$ is the peak absorption
coefficient of the line, $L$ is the absorber path length and
$\Delta\nu$ is the full width at half maximum of the soliton
spectrum. Substituting the actual values of the setup in Fig.
\ref{fig:CrZnSe} ($\beta_{2}=-820\pm40$ fs$^{2}$, $\Delta\nu=$113
cm$^{-1}$= 3.39 THz, round-trip air path length $L=149$ cm, relative
humidity $50\pm1$\% at $21\pm0.5$ $^\circ$C) and taking e.g. the
line at 4088 cm$^{-1}$ (122.56 THz, marked with an
asterisk), we obtain $|\epsilon/q|=0.33\pm0.02$ for the maximum
modulation, which is in perfect agreement with the observed value of
31.5\% (Fig. \ref{fig:CrZnSe}b, black line). The agreement is
remarkably accurate given the less than optimal resolution of the
spectrometer (0.25 cm$^{-1}$) and significant third-order dispersion
of about +10$^{4}$ fs$^{3}$ that was not accounted for in the
presented analysis.

It is important to notice, that the expression (\ref{eq:AmpFinal})
includes only the externally observable soliton bandwidth and relatively
stable dispersion parameter. The alignment-sensitive values
like saturated losses $\sigma$, nonlinearity $\gamma$, nonlinearity
saturation parameter $\kappa$, etc. which are in practice not known
with sufficient accuracy, are all accounted for by the self-consistent
soliton parameters.

Another important point is the fact, that the signal amplitude
$2|\epsilon/q|$ can be much bigger than that from conventional
absorption spectroscopy $\chi_{l}L$ of the cell with the same
length. The signal enhancement factor can be controlled by the pulse
parameters and it exceeds an order of magnitude for the presented
case ($\chi_{l}L=5\%$ for selected line and a single-pass cell of a
resonator size). For additional sensitivity improvement one can
apply the well-developed intracavity multi-pass cell technique
\cite{multi}. The expression (\ref{eq:AmpFinal}) suggests, that
ultimate sensitivity can be obtained at the expense of the reduced
bandwidth coverage $\Delta\nu$. In this respect, the presented
technique has the same quadratic dependence of sensitivity on
spectral bandwidth as the conventional intracavity absorption
spectroscopy \cite{ICLAS-gen}.

Further refinement of the presented theory should include demonstration
of its applicability to arbitrarily shaped absorption features. The
superposition property provides a strong argument for such extension,
but it has to be rigorously proven for Doppler- and more general Voigt-shaped
lines, and also for the dense line groups in e.g. Q-branches. It would
be interesting also to extend the theory to the absorber lines at
the soliton wings (Fig. \ref{fig:CrZnSe}a).

With the above issues resolved, the soliton-based spectroscopy may
become a powerful tool for high-resolution, high-sensitivity spectroscopy
and sensing. Possible implementation include soliton propagation in
a gas-filled holey fibers, as well as already presented intracavity
spectroscopy with femtosecond oscillators.
The latter, being a natural frequency comb source, allows direct locking to optical frequency standards,
providing for ultimate resolution and spectral accuracy.

\section{Conclusion}

We have been able to derive an analytical solution to the problem of
an one-dimensional optical dissipative soliton propagating in a
medium with narrowband absorption lines. We predict appearance of
spectral modulation that follows the associated index of refraction
rather than absorption profile. The novel perturbation analysis
technique is based on integral representation in the spectral domain
and is insensitive to the diverging differential terms, inherent to
the Taylor series representation of the narrow spectral lines.

The model is applicable to a conventional soliton propagation and
to a passively modelocked laser with intracavity absorber, the only
difference being the characteristic propagation distance (dispersion
length and cavity round-trip, respectively). In the latter case the
prediction has been confirmed for a case of water vapour absorption
lines in a mid-IR Cr:ZnSe oscillator. The model provides very good
qualitative and quantitative agreement with experimental observations,
opening a way to metrological and spectroscopical applications of
the novel technique, which can provide a significant (order of magnitude
and more) enhancement of the signal over conventional absorption for
the same cell length.

\begin{acknowledgments}
We gratefully acknowledge insightful discussions and experimental advice from N. Picqu\'{e},
G. Guelachvili (CNRS, Univ. Paris-Sud, France), and I. T. Sorokina (NTNU, Norway).
This work has been supported by the Austrian science Fund FWF
(projects 17973 and 20293) and the Austrian-French collaboration
\emph{Amad\'{e}e}.
\end{acknowledgments}

\end{document}